\documentstyle[multicol,psfig,prb,aps]{revtex}

\begin{document}
\draft

\title{Electron spin resonance in high-field critical phase of gapped
spin chains}

\author{A. K. Kolezhuk\protect\cite{perm}}
\address{Institute of Magnetism, National Academy of Sciences and
Ministry of Education of Ukraine
\protect\\ 
36(B) Vernadskii avenue, 03142 Kiev, Ukraine
\protect\\
Institut f\"ur Theoretische Physik, Universit\"at Hannover,
Appelstra{\ss}e 2, 30167 Hannover, Germany}
\author{H.-J. Mikeska}
\address{Institut f\"ur Theoretische Physik, Universit\"at Hannover,
Appelstra{\ss}e 2, 30167 Hannover, Germany}

\date{\today}

\maketitle

\begin{abstract}
Motivated by recent experiments on $\rm
Ni(C_{2}H_{8}N_{2})_{2}Ni(CN)_{4}$ (commonly known as NENC), we study
the electron spin resonance in the critical high-field phase of the
antiferromagnetic $S=1$ chain with strong planar anisotropy and show
that the ESR spectra exhibit several peculiarities in the critical
phase.  Possible relevance of those results for other gapped spin systems
is discussed.

\end{abstract}


\begin{multicols}{2}

\section{Introduction}
\label{sec:intro}

Recently, there has been a growing interest in the
properties of low-dimensional spin systems subject to strong external
magnetic field.
\cite{SakaiTakahashi91,Chaboussant+98a,Chaboussant+98,Honda+98,Honda+99,%
Papanico+97,Orendac+99,Mila98,ChitraGiamarchi97,ElstnerSingh98,FurusakiZhang99,%
GiamarchiTsvelik99} 
In gapped spin systems, when the external
magnetic field $H$ exceeds a critical value $H_{c}$, it closes the gap
and drives the system into a new critical phase with finite
magnetization and gapless excitations. When the field is further
increased, the system may stay in this critical phase up to the
saturation field $H_{s}$, above which the system is in a saturated
ferromagnetic state. Under certain conditions, however, the
excitations in this high-field phase may again acquire a gap, making
the magnetization ``locked'' in some field range; this phenomenon is
known as magnetization plateaus and has been receiving much attention
as well.\cite{Oshikawa+97,Totsuka98,Shiramura+98,Kageyama+99}

Recently, electron spin resonance (ESR) experiments on $\rm
Ni(C_{2}H_{8}N_{2})_{2}Ni(CN)_{4}$ (commonly abbreviated as NENC) in strong
magnetic fields were conducted.\cite{Orendac+99} This compound is
believed to be a realization of the $S=1$ chain with strong planar and
weak in-plane
anisotropy.\cite{NENC} The theory describing ESR response for this system
outside the critical phase (i.e., $H<H_{c}$ or $H>H_{s}$)
was developed by Papanicolaou et al.\cite{Papanico+97} 
However, much of the ESR data of Ref.\
\onlinecite{Orendac+99} belongs presumably to the field range
$H_{c}<H<H_{s}$, i.e. exactly to the region where the theory is lacking,
which makes the interpretation of the data rather difficult. 

Motivated by those experiments, in the present paper we study
theoretically the zero-temperature ESR response in the critical phase
of a planar $S=1$ chain. It is shown that the typical feature of ESR in
the critical phase is the appearance of continua with resonances being
determined by power-law singularities instead of well-defined
quasiparticle peaks.  We predict that a characteristic change of the
slope in the field dependency of the ESR resonance frequency takes
place at the critical field $H=H_{c}$.

The paper is organized in the following way:
 in Sect.\
\ref{sec:model} we introduce the effective Hamiltonian for the planar
chain, Sect.\ \ref{sec:esr} is devoted to the calculation of resonance
frequencies and exponents characterizing the corresponding
singularities, and, finally, in Sect.\ \ref{sec:discuss} we discuss
possible existence of similar features in other gapped one-dimensional
spin systems.

\section{Planar $S=1$ chain: effective model}
\label{sec:model}

We start from the model of strongly anisotropic  antiferromagnetic
$S=1$ chain described by the Hamiltonian  
\begin{eqnarray} 
\label{ham-largeD} 
\widehat{H}&=&J\sum_{n}{\mathbf S}_{n}\cdot{\mathbf S}_{n+1}
+D\sum_{n}(S_{n}^{z})^{2}
\nonumber\\
&+&E\sum_{n}\Big\{(S_{n}^{x})^{2}-(S_{n}^{y})^{2}\Big\} -h\sum_{n} S_{n}^{z},
\end{eqnarray}
where the planar anisotropy $D$ is assumed to be much stronger than
the exchange constant $J$, $E\ll D$ is a weak in-plane anisotropy, 
and $h=g\mu_{B}H$, where $g$ is the Land\'e
factor, $\mu_{B}$ is the Bohr magneton, and $H$ is the external
magnetic field. The in-plane anisotropy should be included since
it spoils $S^{z}$ as good quantum number and thus allows certain
transitions which are otherwise forbidden; generally, the in-plane
anisotropy constant $E$ may be comparable to $J$.

Let us consider first the noninteracting case $J=0$, then
(\ref{ham-largeD}) amounts to a single-ion problem. The spectrum of a
single ion consists of three states, whose wave functions and energies 
read as follows:
\begin{eqnarray} 
\label{si} 
&& |v\rangle= |0\rangle, \qquad\qquad\qquad\qquad
e_{v}=0,\nonumber\\
&& |a\rangle= \cos\alpha |+\rangle -\sin\alpha |-\rangle, \quad
e_{a}=D-\widetilde{h}\nonumber\\
&& |b\rangle= \sin\alpha |+\rangle +\cos\alpha |-\rangle, \quad
e_{b}=D+\widetilde{h}\nonumber\\
&&\widetilde{h}\equiv
(E^{2}+h^{2})^{1/2},\qquad \sin\alpha\equiv {1\over\sqrt 2}
(1-{h/\widetilde{h}})^{1/2},
\end{eqnarray}
where $|0\rangle$, $|\pm\rangle$ denote the spin-1 states.
For $J\ll D$ and weak field $h<h_{c}$, the ground state of the model
(\ref{ham-largeD}) can be described, to a good approximation, as a
direct product of $|v\rangle$ states of separate magnetic ions.  The
elementary excitations are propagating $|a\rangle$ and $|b\rangle$
states, the spectrum has a gap, and the lowest excitation is a
degenerate doublet. This doublet gets split off by the external field,
$|a\rangle$ state coming down with the field, and $|b\rangle$ states
going up. The low-temperature ESR response for $h<h_{c}$ is determined
by the transitions  from the ground state of the type $|v\rangle \to
|a\rangle$ and $|v\rangle\to|b\rangle$.\cite{Papanico+97}  

When the field $h$ exceeds the critical value
\[
h_{c}=\{(D-2J)^{2}-E^{2} \}^{1/2},
\]
 the system enters critical
phase with a finite density $M$ of $|a\rangle$ states,  and a new type of
ground state transitions, namely of the $|a\rangle\to|b\rangle$ type,
becomes possible (note that this transition
is allowed only in presence of the finite in-plane anisotropy
$E$). The ``old'' types $|v\rangle \leftrightarrow
|a\rangle$ and $|v\rangle\to|b\rangle$ still remain possible. As the
density $M$ increases from $0$ to $1$, the $|v\rangle\to|b\rangle$ signal
should become weaker, while the intensity of the $|a\rangle\to|b\rangle$
transition should increase. Above the saturation field $h_{s}$,
determined by the equation
\[
\widetilde{h}=D+2J\big\{1+(h/\widetilde{h})^{2}\big\},
\]
the density of $a$-particles is equal to one.

In order to describe this picture quantitatively, we first introduce
the hardcore boson operators  $a_{n}^{\dag}$, $b_{n}^{\dag}$ creating
respectively the $|a\rangle$ and $|b\rangle$ states from the vacuum
state $|v\rangle$.  Not more than one boson is allowed to be present
on any site, which defines the set of physical states.  The effective
Hamiltonian in terms of hardcore bosons can be written as
$\widehat{H}_{\rm eff}= {\cal P}({\cal H}_{0}+{\cal H}_{\rm int}){\cal
P}$, where
${\cal P}$ is the projector onto the set of physical states. The
``unperturbed'' Hamiltonian ${\cal H}_{0}$ has the following form:
\begin{eqnarray} 
\label{h0} 
{\cal H}_{0}&=&
\sum_{n}(\Delta_{a}a_{n}^{\dag}a_{n}
+\Delta_{b}b_{n}^{\dag}b_{n})\nonumber\\
&+&\sum_{n} (t_{a}\, a_{n}^{\dag}a_{n+1} +t_{b}\,b_{n}^{\dag}b_{n+1}
+\mbox{h.c.}),
\end{eqnarray}
where the self-energies are $\Delta_{a,b}=e_{a,b}$ and the hopping
amplitudes $t_{a}=t_{b}\equiv t=J$. 
The interaction part ${\cal H}_{\rm int}$ looks as follows:
\begin{eqnarray}
\label{hint} 
{\cal H}_{\rm int}&=&
 U_{aa}\sum_{n} (a_{n}^{\dag}a_{n})(a_{n+1}^{\dag}a_{n+1})\\
&+& U_{ab}\sum_{n}\big\{ (a_{n}^{\dag}a_{n})(b_{n+1}^{\dag}b_{n+1}) +
(b_{n}^{\dag}b_{n})(a_{n+1}^{\dag}a_{n+1}) \big\},\nonumber
\end{eqnarray}
where the interaction constants are given by
$U_{aa}=-U_{ab}=J(h/\widetilde{h})^{2}$.

Our goal is to take into account the effects arising in the first
order in the coupling $J$, and therefore  we have
neglected the pair creation term
\[
\Phi_{ab}\sum_{n}(a_{n}b_{n+1}+b_{n}a_{n+1}+\mbox{h.c.}),
\]
with $\Phi_{ab}=J(h/\widetilde{h})$, whose contribution to any energy
level arises only in the second order in $J$. This simplifies the
problem considerably, making the total numbers $N_{a}$ and $N_{b}$ of
$a$ and $b$ particles good quantum numbers. At the same level of
approximation, one can neglect the processes of exchanging
the positions of neighboring $a$ and $b$ particles, described by the
term 
\[
t_{ab}\sum_{n} \big\{ 
b_{n}^{\dag}a_{n+1}^{\dag}a_{n}b_{n+1}
+\mbox{h.c.}\big\},
\]
since the  `interparticle' hopping amplitude
$t_{ab}=J(E/\widetilde{h})^{2}$ is very small 
in the critical region, $E/\widetilde{h}\sim E/D\ll 1$.  Further,
we will be interested in configurations with at most one $b$ boson, so
that the interaction between $b$ particles can be also safely
neglected.

Let us first consider the ``unperturbed'' Hamiltonian $\cal P
H_{\rm 0}P$ (note that 
the physics described by this Hamiltonian is nevertheless
nontrivial due to the single occupancy constraint).

In absence of $b$-particles ($N_{b}=0$) the spectrum of the problem
can be in principle obtained by mapping to
noninteracting fermions with the help
of the well-known Jordan-Wigner transformation, but it is more
convenient to stick to the hardcore boson language. From the
Bethe-ansatz solution \cite{Gaudin} one knows that the energy spectrum is given by
\begin{equation} 
\label{e-ba} 
E(\{k_{l}\})=\sum_{l} (\Delta_{a}+\cos k_{l}),\quad k_{l}=\pi+{2\pi\over
L}I_{l},
\end{equation}
where the numbers $I_{l}$ are all different and respectively
half-integer (integer) if the total number of particles $N=N_{a}$ is
even (odd), and $L$ is the number of sites which will be assumed even
for convenience. The total momentum $P$ of a state defined by a
certain choice of $N$ different numbers $I_{l}$ is
$P=\sum_{l}k_{l}$. The ground state $|{\rm g.s.}\rangle$ is given by a
symmetrical dense distribution of $I_{l}$ around zero:
\begin{equation}
\label{i-ba}
I_{l}=-{N-1\over2},-{N-3\over2},\ldots, {N-1\over2}\,,
\end{equation}
describing the Fermi sea of particles with momenta in the interval
$[k_{F},2\pi-k_{F}]$, where we shall define
\begin{equation} 
\label{kF} 
k_{F}=\pi(1 - N/L)\,.
\end{equation}
The total momentum of the ground state configuration is $P=0\,
(\mbox{mod}\, 2\pi)$ if $N$
is even, and $P=\pi$ if $N$ is odd.  

Since the hopping amplitudes for
$a$- and $b$-particles are equal, it is easy to realize that the above
picture of the distribution of wave vectors remains true when one has
the total number of particles $N=N_{a}+N_{b}$, $N_{a}$ of them being
of the $a$ type and $N_{b}$ of the $b$ type: they form a single
``large'' Fermi sea.

\section{ESR transitions}
\label{sec:esr}

We would like to study the ESR transitions from the ground state,
which survive in the low-temperature limit. The ESR
intensity $I(\omega)$ 
for the ground state transitions is determined by the
formula\cite{Slichter78}
\begin{equation} 
\label{esr} 
I(\omega)\propto  \omega \sum_{f}\Big|\sum_{\alpha} h_{r}^{\alpha}
\langle f|S^{\alpha}_{\rm tot}|{\rm
g.s.}\rangle\Big|^{2}\delta(E_{f}-E_{\rm g.s.}-\omega),
\end{equation}
where $h_{r}^{\alpha}$ are the components of the radio frequency field,
and $f$ labels all possible final states.  

\subsection{$|v\rangle\mapsto |b\rangle$ transitions}

Let us first consider the $|v\rangle\mapsto |b\rangle$ process with
$\Delta N_{a}=0$ and $\Delta N_{b}=1$, as the most interesting one
(note that it is allowed even in absence of the in-plane anisotropy
$E$). This process is determined by the operator $S^{-}_{\rm tot}$,
which is in this case equivalent to $\sqrt{1-h/\widetilde{h}}
\sum_{n}b^{\dag}_{n}$.  Let us
assume for convenience that the initial number of particles $N=N_{a}$
is even, so that $N=2N_{0}$.  Changing the total number of particles
by one causes, according to (\ref{i-ba}), the rearrangement of the
allowed wave vectors. This leads to the vanishingly small transition
matrix elements in (\ref{esr}) (orthogonality
catastrophe\cite{Anderson67}), however it is compensated by a
diverging number of states $|f\rangle$ having nearly the same energy.
The problem of calculating the ESR response in this case amounts to
that of calculating the spectral properties of a mobile hardcore
impurity suddenly created in a hardcore boson system, the impurity
having the same mass as the other particles, and the dispersion law
being of the type $\Delta_{a,b}+2t\cos k$. A similar problem was
studied in Refs.\
\onlinecite{Ogawa+92,CastellaZotos93,LiMa95,Tsukamoto+98} and is
closely related to the so-called Fermi edge singularities in the
photoemission/adsorption spectra (see, e.g., Ref.\
\onlinecite{OhtakaTanabe90} and references therein).

The excited wave function $|f\rangle$ can be generally
represented in the following form: 
\begin{equation} 
\label{fin} 
|f\rangle= \sum_{x_{0}'} e^{iP'x_{0}'}
 \sum_{y_{1}'\ldots y_{N}'} \Phi(y_{1}',\ldots,y_{N}') | x_{0}';\; 
y_{1}'\ldots y_{N}'\rangle ,
\end{equation}
where $x_{0}'$ denotes the position of the $b$ particle and
$x_{i}'=x_{0}'+y_{i}'$, $i=1,\ldots,N$ denote the positions of the
$a$-particles (so that $y_{i}'$ are their coordinates relative to the
$b$-particle), and $P'$ is the total momentum of the excited
state. 

Following Castella and Zotos, \cite{CastellaZotos93} one can
write the reduced wave function $\Phi$ in the determinantal form:
\begin{eqnarray} 
\label{CZ} 
&&\Phi(y_{1}',\ldots,y_{N}')=(-)^{\Sigma'}
{1\over \sqrt {LN!}}\det \{\varphi_{i}(y'_{j})\},\nonumber\\
&& \varphi_{l}(y)=A_{l}\Big\{ e^{i(k_{l}'y+\delta_{l})} 
-{1\over N}\sum_{n=1}^{N}e^{i(k_{n}'y+\delta_{n})} \Big\},
\end{eqnarray}
where $A_{l}$ is the normalization factor, and the phase shift
$\delta_{l}\equiv\delta=-\pi/2$ is  independent of the state label $l$
in our case of noninteracting
hardcore particles.  The factor
$(-)^{\Sigma'}=\pm1$ is the sign of the permutation $\Big(
\displaystyle {y_{1}'\atop y_{i_{1}}'} {y_{2}' \atop y_{i_{2}}'}
{\cdots\atop\cdots} {y_{N}' \atop y_{i_{N}}' }\Big)$, where
$y_{i_{1}}'< y_{i_{2}}' < \cdots <y_{i_{N}}'$, which ensures that the
total wave function is symmetric under the permutation of any two
particles.  The allowed values for the wave vectors $k_{l}'$ of the
excited state are given by (\ref{e-ba}) with integer $I_{l}$. The
energy of this state is given by
\begin{equation} 
\label{e-fin}
E_{f}=\sum_{l}(\Delta_{a}+2t\cos k_{l}')+\Delta_{b}+2t\cos\lambda, 
\end{equation}
where $\lambda$  is defined
by
\begin{equation} 
\label{lambda} 
\lambda=P'-\sum_{l}k_{l}'
\end{equation}
and plays the role of the  momentum of the $b$ particle.

It is easy to see that the wave function $b^{\dag}_{x_{0}}|{\rm
g.s.}\rangle$, obtained from the ground state by creating a
$b$-particle at site $x_{0}$, can be
written in a similar determinantal form:
\begin{equation} 
\label{ini}
b^{\dag}_{x_{0}}|{\rm g.s.}\rangle = e^{iPx_{0}}\sum_{y_{1}\ldots
y_{N}} f(y_{1},\ldots,y_{N}) | x_{0};\; y_{1}\ldots y_{N}\rangle ,
\end{equation} 
where $y_{i}=x_{i}-x_{0}$ are again relative coordinates of
$a$-particles, $P=\sum_{l}k_{l}$ is the total momentum, and the
reduced wave function
\begin{equation} 
\label{plane} 
f(y_{1},\ldots,y_{N})=(-)^{\Sigma}{1\over \sqrt {N!}}\det \{\psi_{i}(y_{j}) \},\nonumber\\
\end{equation}
is built from simple plane waves
$\psi_{l}(y)=(1/\sqrt{L})e^{ik_{l}y}$, $(-)^{\Sigma}$ is the sign
of the permutation $\Big(\displaystyle{y_{1}\atop y_{i_{1}}} {y_{2} \atop
y_{i_{2}}} {\cdots\atop\cdots} {y_{N} \atop y_{i_{N}} }\Big)$, where
$y_{i_{1}}< y_{i_{2}} < \cdots <y_{i_{N}}$.

Now, one can see that the matrix element entering (\ref{esr}) takes
the form
\begin{eqnarray}
\label{fi}
&&\langle f| \sum_{n}b_{n}^{\dag} |{\rm
g.s.}\rangle=\sqrt{L}\,\delta(P,P') M_{fi},\nonumber\\
&&M_{fi}= \sum_{y_{1}y_{2}\cdots y_{N}} \Phi^{*}(y_{1},y_{2},\ldots,y_{N})\,
 f(y_{1},y_{2},\ldots,y_{N}),
\end{eqnarray}
where $\delta(P,P')$ is the Kronecker symbol telling us the well-known fact
that the total momentum should not change during the ESR transition,
$P=P'\, (\mbox{mod}\, 2\pi)$. The reduced matrix element $M_{fi}$ can
be represented as a determinant of the overlap matrix between `old'
and `new' wave functions,
\begin{equation} 
\label{fi1} 
M_{fi} =\det\{\langle \varphi_{l'}| \psi_{l}\rangle \}.
\end{equation}
As shown in Ref.\ \onlinecite{CastellaZotos93}, the wave functions
$\varphi_{l}(y)$ defined in (\ref{CZ}) 
behave asymptotically as plane waves,
 $\varphi_{l}(y)\mapsto \widetilde{\varphi}_{l}(y)
=(1/\sqrt{L})e^{i(k_{l}'y+\delta_{l})}$, and the overlaps (\ref{fi1})
calculated with $\varphi_{l}$ can be with the accuracy of $O(\ln L/L)$
replaced by those calculated with $\widetilde{\varphi}_{l}$. Then the
matrix element $M_{fi}$ is asymptotically equal to the overlap between
two Slater-type wave functions, one describing the system with the
total momentum $P=\sum_{l}k_{l}$ and the other corresponding to the
system with the momentum $Q=\sum_{l}k_{l}'\equiv P'-\lambda$. It is
thus clear that $M_{fi}$ can be nonzero only if $P=Q$. This gives us
the complete set of selection rules as
\begin{equation} 
\label{selrule} 
P=P',\quad
\lambda=0.
\end{equation}

The ground state configuration is given by the following distribution
of the wave vectors:
\[
k_{l}=\pi\pm {2\pi\over L}(l+{1\over2}),\quad l=0,\ldots, (N_{0}-1)
\]
and has the total momentum $P=2\pi N_{0}$, while in the excited state the
numbers $I_{l}$ are integer, and the lowest energy configuration built
according to (\ref{i-ba}) would have the total momentum $\pi(2N_{0}+1)$.  In
order to get the momentum change $\Delta P$ back to zero, one has to
introduce some excitation.  In the
simplest way this can be achieved by creating a hole at $k=\pi$, i.e.,
one gets the symmetric excited state configuration $|f_{s}\rangle$
given by
\begin{equation} 
\label{config1} 
k_{n}'=\pi\pm {2\pi \over L}(n+1),\quad n=0,\ldots,N_{0}-1.
\end{equation}
The overlap $M_{fi}^{s}$ can be easily
calculated by means of passing to the even/odd states (i.e., to sines
and cosines of $k_{l}y$ and $(k_{l}'y+\delta)$, respectively). The
determinant then factorizes into a product of two (in our case equal)
Cauchy-type determinants $M^{(\pm)}$, 
\[
M^{(\pm)}=\det\Big\{ {1\over \pi(n-n')-\delta} \Big\} \,.
\]
Determinants of this type were calculated by Anderson
\cite{Anderson67} and shown to be algebraically vanishing in the
thermodynamic limit, $M^{(\pm)}\propto
L^{-\beta_{\pm}/2}$, with
the exponents
$\beta_{\pm}=(\delta/\pi)^{2}=1/4$, so that 
\begin{equation} 
\label{sym} 
|M_{fi}^{s}|^{2}\propto
L^{-\beta_{s}}
\end{equation}
 with the orthogonality catastrophe (OC) exponent
$\beta_{s}=\beta_{+}+\beta_{-}=1/2$. This exponent can be also
calculated in a different way, using the results of boundary conformal
field theory (BCFT).\cite{Affleck+94-97} For this purpose it is necessary to
calculate the energy difference $\Delta E_{f}$ between the ground
state and the excited state $|f\rangle$, including the $1/L$
corrections. Then in case of \emph{open boundary conditions} the OC
exponent $\beta$, according to BCFT, can be obtained as
\begin{equation} 
\label{bcft} 
\beta={2L \widetilde{\Delta E_{f}}\over \pi v_{F} }\equiv
{\widetilde{2\Delta E_{f}} \over \Delta E_{\rm min}}.
\end{equation}
Here $v_{F}=2t\sin k_{F}$ is the Fermi velocity, so that $\Delta
E_{\rm min}=\pi v_{F}/L$ is the lowest possible excitation energy, and
$\widetilde{\Delta E_{f}}$ is the $O(1/L)$ part of $\Delta E_{f}$
(i.e., with the bulk contribution subtracted). In this last form this
formula should be also valid for the \emph{periodic boundary
conditions}, then $\Delta E_{\rm min}$ should be replaced by $2\pi
v_{F}/L$. It is easy to calculate $\Delta E_{f}$ for the symmetric
configuration $|f_{s}\rangle$: one gets $\Delta
E_{f}^{s}=\Delta_{b}+2t+\Delta E'_{s}$, where
\begin{eqnarray} 
\label{de-sym}
\Delta E'_{s}&=&
4t\sum_{l=0}^{N_{0}-1}\Big\{\cos {2\pi\over
L}(l+{1\over2}) -\cos {2\pi \over L}(l+1) \Big\} \nonumber\\
&=&2t(1+\cos k_{F})
+ t{\pi\over L}\sin k_{F}+O(1/L^{2}).
\end{eqnarray} 
Applying (\ref{bcft}), one gets the correct value $\beta=1/2$ which
coincides with that obtained through a traditional way by calculating
determinants. 

The symmetric excited state configuration with $Q=P$ is obviously
\emph{not} the configuration with the lowest energy.  The symmetric
configuration \emph{is}, however, the configuration with the highest
overlap with the ground state.  Indeed, there are configurations with
asymmetric distribution of momenta around $k=\pi$, which nevertheless
satisfy the selection rules, e.g., the following one which we will
denote as $|f_{a}(n)\rangle$:
\begin{equation} 
\label{asym} 
k_{l}'=\pi+{2\pi\over L}l,\quad l=-(N_{0}-n),\ldots,(N_{0}+n-1).
\end{equation} 
If we require that
$Q=P+2\pi\Lambda$, where $\Lambda$ is an integer, the following
condition on $k_{F}$ is obtained:
\begin{equation} 
\label{kF-asym}
k_{F}=\pi {2m-1\over 2n-1},\quad m\equiv n-\Lambda, 
\end{equation}
which gives the dense set of allowed values of $k_{F}$, at which the 
configuration of the type (\ref{asym})
will satisfy the selection rules.  
The energy difference from the
ground state energy is  $\Delta E_{f}^{a}(n)=\Delta_{b}+2t+\Delta E'_{a}(n)$,
with 
\begin{equation} 
\label{de-asym} 
\Delta E_{a}'(n)=t(2n-1)^{2}{\pi\over L}\sin k_{f} +O(1/L^{2}).
\end{equation}
The corresponding OC exponent $\beta_{a}(n)$, according to (\ref{bcft}), is
given by 
\begin{equation} 
\label{beta-asym} 
\beta_{a}(n)={1\over2}(2n-1)^{2}
\end{equation} 
and is always larger than the OC exponent for the symmetric
configuration $\beta_{s}=1/2$ (except for $n=1$, which would mean
$k_{F}=\pi$, i.e., the vanishing Fermi sea). 

\begin{figure}
\psfig{figure=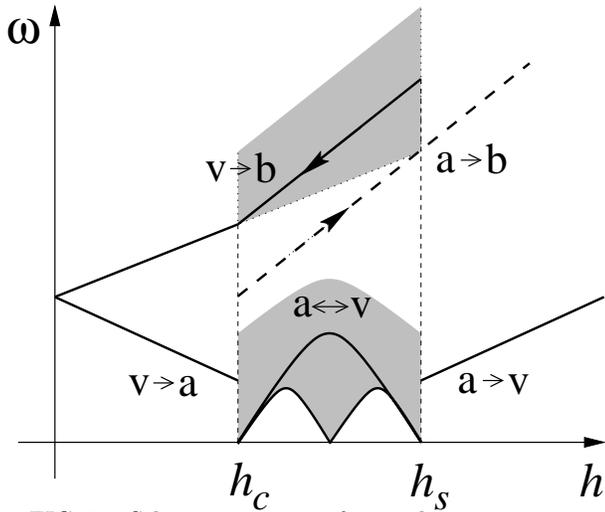,width=80mm}
\caption{\label{fig:lines} Schematic picture of ground state
transitions in the ESR spectrum of the
planar $S=1$ chain as functions of the field in the critical
region. Dashed line corresponds to the quasiparticle peak determined
by $a\to b$ processes which are
forbidden in absence of the in-plane anisotropy $E$. Filled areas show
the continua, and solid lines within the continua
indicate the position of the singularities. Arrows on the lines
indicate the direction of the intensity increase when varying the
magnetic field.}
\end{figure}

The main contribution into the ESR response is thus coming from the
states whose energy is close to that of the symmetric
configuration. The summation over ``shake-up'' (i.e., particle-hole)
excitations around the symmetric configuration leads to the
singularity in the ESR intensity $I(\omega)$ of the form 
\cite{OhtakaTanabe90}
\begin{equation} 
\label{FES} 
I(\omega)\propto 1/(\omega-\omega_{0})^{\alpha},\quad \omega>\omega_{0}
\end{equation}
with $\alpha=\alpha_{s}=1-\beta_{s}=1/2$, and the threshold frequency
$\omega_{0}=\Delta E_{f}^{s}=\Delta_{b}+4t+2t\cos k_{F}$.
It is remarkable that the  contribution from
the asymmetric configuration displays no singularity, since the
corresponding OC exponent $\beta_{a}>1$, so that $\alpha_{a}<0$; this
contribution yields just some scattered intensity at frequencies
$\omega\geq \Delta_{b}+2t$, see Fig.\ \ref{fig:lines}. 
The integrated intensity of $|v\rangle\to|a\rangle$ contribution is
proportional to  the total number $L(1-M)$ of empty sites, and thus
decreases with the magnetic field; this contribution does not exist
above the saturation field $H_{s}$.

Taking into account the definition of $k_{F}$, namely
$\Delta_{a}+2t\cos k_{F}=0$, the formula for threshold $\omega_{0}$
can be rewritten as
\begin{equation} 
\label{edge} 
\omega_{0}(v\to b)=\Delta_{b}-\Delta_{a}+4t
\end{equation}
One may notice the following remarkable feature: for weak magnetic
fields below $h_{c}$ the $|v\rangle \mapsto |b\rangle$ process leads
to a quasiparticle  peak 
in the ESR response at the frequency $\omega_{0}'=\Delta_{b}+2t$, so
that \emph{below the critical field} the slope $d\omega_{0}'/dh$ of
the ESR line as a function of the field $h$ is approximately $1$ (for
$E\ll D)$. For $h>h_{c}$ the same process yields the Fermi-edge-type
singularity (\ref{FES}) in the ESR response, with the edge frequency
$\omega_{0}$ being given by (\ref{edge}). If one makes a reasonable
assumption, that experimentally observed ESR line for $h>h_{c}$
follows the behavior of the edge singularity, 
we get for the slope of the ESR line
\emph{above the critical field} the value $d\omega_{0}/dh\approx
2$. Thus, the slope of the experimentally observed line should change
abruptly at the critical field $h=h_{c}$, as shown schematically in
Fig.\ \ref{fig:lines}. It is worthwhile to note that 
such a behavior is reminiscent of the picture
experimentally seen in NENC \cite{Orendac+99} in the intermediate
field regime.

Though taking into account the effects of temperature is beyond the
scope of the present paper, we would like to remark that the singular
contribution of $|v\rangle\to|b\rangle$ processes to the ESR response 
should be enhanced at finite temperature, due to the presence of a
finite number of holes in the ground state. The density of states has
a singularity at the bottom of the Fermi sea, so that the main
contribution will come at the same frequency (\ref{edge}). This could
be important for interpreting the results of experiments on NENC,
\cite{Orendac+99} where the temperature was of the order of $J$.

If the interaction terms (\ref{hint}) are taken
into account, the corresponding effective model cannot be solved
exactly any more, and we cannot calculate the OC exponents in this
case. 
However, one can treat the effect of interactions in
a sort of the ``mean-field'' approximation, i.e., replacing simply 
\begin{eqnarray} 
\label{mean-field} 
&&\Delta_{a}\mapsto \Delta_{a}(M)=\Delta_{a}+U_{aa}M,\nonumber\\ 
&&\Delta_{b}\mapsto \Delta_{b}(M)=\Delta_{b}+U_{ab}M,
\end{eqnarray}
where the particle density  $M=1-k_{F}/\pi$ has to be calculated
self-consistently from the equation
\begin{equation} 
\label{sc} 
\Delta_{a}(M)+2t\cos k_{F}=0.
\end{equation}
One can check that such an approximation, though being crude, delivers
correct values for the critical fields $H_{c}$ and $H_{s}$. Actually,
in absence of $b$-particles one can show that the effect of $U_{aa}$
is not only to change the self-energy $\Delta_{a}$, but also to
renormalize the hopping amplitude $t_{a}$ (in the first order in
$U_{aa}$ one gets $t_{a}\mapsto t_{a}+(U_{aa}/\pi)\sin k_{F}$), and
both $t_{a}$ and $U_{aa}$ are of the order of $J$, so this correction
is not negligible.  According to the numerical
studies,\cite{Castella96} the OC exponent $\beta$ is rather sensitive
to the ratio of hopping amplitudes $r=t_{b}/t_{a}$ (the behavior of
$\beta(r)$ is approximately linear around $r=1$), so that the
corresponding OC exponents will certainly change compared to the
noninteracting (hardcore) case, and they will become
$k_{F}$-dependent.  One may nevertheless expect that at the
\emph{qualitative} level our picture of transitions remains true,
particularly, the effect of changing the slope of ESR lines at
$H=H_{c}$ should survive.

\subsection{$|a\rangle\mapsto |b\rangle$ transitions}

The $|a\rangle\mapsto |b\rangle$ processes are possible only in
presence of the finite in-plane anisotropy $E$, which mixes
$|-\rangle$ and $|+\rangle$ states into $|a\rangle$ and $|b\rangle$,
respectively. Those transitions are determined by the operator
$S^{z}_{\rm tot}$, which is in this case equivalent to
$(E/\widetilde{h})\sum_{n}b_{n}^{\dag}a_{n}$. One can easily see that
now the wave functions $\sum_{n}b_{n}^{\dag}a_{n} |{\rm g.s}\rangle$
and $|f\rangle$ can be both represented in the form (\ref{fin}),
(\ref{CZ}) with $N\to (N-1)$. The total momentum of the ground state
can be represented as $P=\sum_{i=1}^{N-1}\widetilde{k}_{i}
+\widetilde{\lambda}$, and the momentum of the excited state has a
similar form, $P'=\sum_{i=1}^{N-1}k'_{i} +\lambda$. Through the same
line of arguments as before one gets the selection rules
\[
P=P',\quad \lambda=\widetilde{\lambda}.
\]
Since the number of particles was not changed during the transition, 
the allowed values of wave vectors $k'_{i}$ are now the same as 
$\widetilde{k}_{i}$ and are given by $\pi+(2\pi/L)I_{i}$ with
half-integer $I_{i}$. Thus, there is \emph{no orthogonality
catastrophe} in this case, and one gets a well-defined \emph{quasiparticle
peak} at the frequency
\begin{equation} 
\label{atob} 
\omega_{a\to b}=\Delta_{b}-\Delta_{a}.
\end{equation}
One can visualize this process as simply replacing one of the $a$-particles
in the Fermi sea of the ground state by a $b$-particle, without
changing the wave vector  distribution. The corresponding ESR line
goes parallel to the $|v\rangle\to |b\rangle$ line, slightly below it,
as shown schematically in Fig.\ \ref{fig:lines}. The intensity of this
line is proportional to the total number $LM$ of $a$-particles and
increases with the field; this line continues to exist at $H>H_{s}$.  

\subsection{$|v\rangle\leftrightarrow |a\rangle$ transitions}

Those transitions are allowed in absence of the in-plane
anisotropy. The number of particles changes by one, which leads to a
change in the distribution of momenta and indicates absence of the
quasiparticle peak.  Since only two states per site are involved
($N_{b}$ remains zero), those processes can be treated within the
effective spin-$1\over2$ model, similarly to the way it was done for
spin ladders.\cite{Mila98,Totsuka98} If one neglects the small
corrections of the order of $(E/h)^{2}$, this effective model is the
easy-plane XXZ chain in an effective longitudinal magnetic field with
the Hamiltonian\cite{Krohn}
\[
H_{eff}=\sum_{n} \widetilde{J}_{xy}(\widetilde{S}_{n}^{x}\widetilde{S}_{n}^{x} 
+\widetilde{S}_{n}^{y}\widetilde{S}_{n}^{y})
+\widetilde{J}_{z}\widetilde{S}_{n}^{z}\widetilde{S}_{n}^{z} 
-\widetilde{B}\widetilde{S}_{n}^{z}
\]
where $\widetilde{S}^{\mu}$ are the spin operators of the effective
spin-$1\over2$ chain, $J_{xy}=2J$, the easy-plane anisotropy
$J_{z}/J_{xy}=1/2$, and the field $\widetilde{B}=h-J-D$.

Calculating the low-frequency ESR response amounts then to
the knowledge of the dynamical structure factor $S^{+-}(q,\omega)$ of
the effective XXZ chain, $I(\omega)\propto S^{+-}(q=0,\omega)$. The
asymptotic behavior of the correlation functions for the XXZ chain 
is known from the
bosonization results:\cite{LutherPeschel75,FurusakiZhang99}
\begin{eqnarray} 
\label{correl}
&&\langle \widetilde{S}^{+}(x,t)\widetilde{S}^{-}(0,0)\rangle
= {C_{1}\over (x^{2}-v_{F}^{2}t^{2})^{1/(4K)}}
\cos(\pi x)\\
&&\;\;+{C_{2}\over (x^{2}-v_{F}^{2}t^{2})^{K+1/(4K)-1}}
\Big\{ {e^{i(2k_{F}-\pi)x}\over (x-v_{F}t)^{2}} 
+ {e^{-i(2k_{F}-\pi)x}\over (x+v_{F}t)^{2}}\Big\}. \nonumber
\end{eqnarray}
Here $C_{1,2}$ are some constants, 
$v_{F}$ is the Fermi velocity and $K$ is the so-called Luttinger
liquid parameter. For $\widetilde{B}=0$ (which corresponds to
$k_{F}=\pi/2$, or $M=1/2$, or equivalently to the physical field $H$
in the middle of the critical region,
$H={1\over2}(H_{c}+H_{s})$)  and $J_{z}/J_{xy}=1/2$ one has 
\cite{GiamarchiTsvelik99}
\begin{equation} 
\label{vF,K}
v_{F}=3\sqrt{3}/2,\quad K=3/4\,.
\end{equation}
At nonzero $\widetilde{B}$ (i.e., for $M\not=1/2$) the values of
$v_{F}$ and $K$ can be obtained numerically by solving the Bethe
ansatz equations;\cite{Korepin+} the corresponding solutions are
presented in Fig.\ 2 of Ref.\ \onlinecite{GiamarchiTsvelik99}. For
$k_{F}\to \pi$ ($0$), which corresponds to $H\to H_{c}$ ($H_{s}$),
respectively, i.e., when  the number of particles (holes) becomes
small, $K$ tends to $1/2$ and $v_{F}$ to zero. 

Fourier-transforming  (\ref{correl}),
one gets the $q=0$ dynamic
structure factor $S^{+-}(q=0,\omega)$ 
having edge-type  singularities at the frequencies 
\begin{equation} 
\label{vtoa} 
\omega_{0}(Q)=v_{f}Q
\end{equation}
where $Q=\pi$
and $Q=2k_{F}-\pi$
for the contributions of the first and second term  in (\ref{correl}),
respectively. 
The singularities are of power-law type, with the exponent
being determined by the Luttinger liquid parameter $K$:
\begin{eqnarray}
\label{expvtoa}
&&I(\omega)\propto {1\over 
\big[\omega-\omega_{0}(Q)\big]^{\eta(Q)}},\nonumber\\
&&\eta(0)=1-{1\over 4K},\quad \eta(2k_{F}-\pi)=2-K-{1\over 4K}.
\end{eqnarray}
Thus, two lines corresponding to the singularities in the
$|v\rangle\leftrightarrow |a\rangle$ processes at
$\omega=\omega_{0}(\pi)$ and $\omega=\omega_{0}(2k_{F}-\pi)$
should be visible in the ESR spectrum, as shown schematically in Fig.\
\ref{fig:lines}. Inclusion of the finite temperature would lead to
damping of the singularities.\cite{ChitraGiamarchi97}

\section{Discussion}
\label{sec:discuss}

In the model of planar $S=1$ chain considered in the previous section
we have seen several features of the ESR spectrum in the high-field
critical phase: appearance of low-lying and high-lying continua with
power-law singularities, counter-intuitive change of the slope of ESR
lines at the critical field $H=H_{c}$, etc. One may ask oneself
whether similar features can be present also in other one-dimensional
spin systems, e.g., strongly coupled $S={1\over2}$ ladders like
$\rm Cu_{2}(C_{5}H_{12}N_{2})_{2}Cl_{4}$ (usually abbreviated
CuHpCl\cite{Chaboussant+98}) and $\rm (C_{5}H_{12}N_{2})_{2}CuBr_{4}$
(abbreviated BPCB\cite{Watson+00}), or $S=1$ Haldane chain $\rm
Ni(C_{5}H_{14}N_{2})_{2}N_{3}(PF_{6})$ (known as NDMAP
\cite{Honda+98,Honda+99}).

As far as the model of $S={1\over2}$ ladder with strong rung
interaction $J_{R}$ and weak leg coupling $J_{L}\ll J_{R}$ is
concerned, one can construct the effective model exactly in the same
way as for the large-$D$ $S=1$ chain. The role of vacuum state
$|v\rangle$ will be played by singlet state $|s\rangle$ on a single
rung, and one will have three types of ``particles'' corresponding to
the three triplet states $|t_{\pm}\rangle$, $|t_{0}\rangle$. Applying
a strong magnetic field will lead to closing the gap and to a finite
density of $|t_{+}\rangle$ states in the ground state in the critical
phase.  However, in absence of any additional couplings (e.g.,
Dzyaloshinskii-Moriya interaction or non-uniaxial anisotropies) the
only possible direct transitions from the ground state are determined
by the processes of the $|t_{+}\rangle\to |t_{0}\rangle$ type, which,
similarly to the $|a\rangle\to |b\rangle$ processes considered in the
previous section, yield a quasiparticle peak with the resonance
frequency $\omega_{+0}=h$.  Only in presence of such additional interactions
providing finite admixture of triplets in $|v\rangle$ one could see
$|v\rangle\to|t_{0}$ and $|v\rangle\to |t_{-}\rangle$ lines, which
should then exhibit the change of slope at $H=H_{c}$ with the
simultaneous appearance of the continuum above them.  
Similar arguments apply also to the $S=1$ Haldane chains: generally,
in order to observe the ``interesting'' lines, one needs to have some
perturbations allowing the direct transitions from the ground state at
$H=0$. 

In this sense, the planar $S=1$ chain is a remarkable model, where the
features like the slope change or the low-lying continuum can be
observed ``generically'', without appealing to the existence of any
additional interactions. However, one may in principle hope that
similar effects could be observed in other one-dimensional spin
systems as well.  There are, for example, experimental indications
\cite{Ohta+00} that such additional terms are really present in
CuHpCl.

In recent experiment \cite{Honda+99} on the $S=1$ Haldane chain
compound NDMAP it was observed, that one of the ESR branches was just
continuing into the critical region, without any noticeable features
at $H=H_{c}$. It is worth noting that this feature closely resembles
our conclusions for the $|a\rangle\to|b\rangle$ processes in the
planar chain: if one denotes the (field-dependent) magnon gaps for
NDMAP as $\Delta_{\pm}$ and $\Delta_{0}$, then at $H<H_{c}$ the
above-mentioned ESR branch corresponds to the thermally excited
transition within the magnon triplet, with the frequency 
$\omega=\Delta_{0}-\Delta_{+}$. At $H>H_{c}$ there is a finite
density of $\Delta_{+}$-magnons in the ground state, and, if we
exploit the analogy with our picture of $|a\rangle\to|b\rangle$
processes for the planar chain, one can expect the presence of a
quasiparticle peak at exactly the same frequency, in agreement with
the experiment. 

Finally, some words of caution are here in order. In the present
paper, we have studied only the \emph{ground state transitions}, in
other words the ESR response at zero temperature, and only in purely
one-dimensional (1D) model. When interpreting the experimental data,
one should have in mind that, because of the gapless nature of the
ground state in the critical phase of the 1D system, the effects of
temperature become important, as well as those of weak 3D coupling;
particularly, the system should be 3D ordered under certain critical
temperature. Thus, the results presented here should only be taken as
a guide displaying features of the purely 1D behavior. Further work is
required to analyze the possibility of observation of those phenomena
in realistic systems.

\section*{Acknowledgements}

AK gratefully acknowledges the hospitality of Hannover Institute for
Theoretical Physics.  This work was supported by the German Federal
Ministry for Research and Technology (BMBFT) under the contract
03MI5HAN5 and by the Volkswagen Foundation through the grant I/75895.

\end{multicols}

\end{document}